\documentclass[groupedaddress,pre,twocolumn]{revtex4}
\usepackage{amssymb}
\usepackage{bm}
\usepackage{amsmath}
\usepackage{graphicx}
\usepackage{color}

\begin{document}

\title{Metadynamics with adaptive Gaussians}
\author{Davide Branduardi}
\affiliation{Theoretical Molecular Biophysics Group, Max Planck Institute for Biophysics, Max-von-Laue strasse 5, 60438, Frankfurt am Main, Germany}
\author{Giovanni Bussi}
\email{bussi@sissa.it}
\affiliation{SISSA - Scuola Internazionale Superiore di Studi Avanzati, via Bonomea 265, 34136, Trieste, Italy}
\author{Michele Parrinello}
\affiliation{
Department of Chemistry and Applied Biosciences, ETH Zurich, Via G.~Buffi 13, 6900, Lugano, Switzerland}
\affiliation{
Facolt\`a{} di Informatica, Istituto di Scienze Computazionali,
Universit\`a{} della Svizzera Italiana, Via G.~Buffi 13, 6900, Lugano, Switzerland
}



\begin{abstract}
Metadynamics is an established sampling method aimed at reconstructing the free-energy surface
 relative to a set of appropriately chosen collective variables.
In standard metadynamics the free-energy surface is filled by the addition of Gaussian potentials 
of pre-assigned and typically diagonal covariance.  Asymptotically the free-energy surface is  proportional 
to the bias deposited. Here we consider the possibility of using Gaussians
whose variance is adjusted on the fly to the local properties of the free-energy surface. 
We suggest two different prescriptions: one is based on the local diffusivity and the other on the local geometrical properties.
We further examine the problem of extracting the free-energy surface when using adaptive Gaussians. 
We show that the standard relation between the  bias and the free energy does not hold. 
In the limit of narrow Gaussians an explicit correction can be evaluated. 
In the general case we propose to use instead a relation between bias and free energy borrowed from umbrella sampling. 
This relation holds for all kinds of incrementally deposited bias.  
 We illustrate on the case of alanine dipeptide the advantage of using adaptive Gaussians in conjunction with the new free-energy estimator both in terms of accuracy and speed of convergence. 
\end{abstract}
\maketitle

\section{Introduction}

The problem of sampling complex energy surfaces, characterized by metastable states 
separated by large energy barriers, has recently received considerable attention. 
A list that is by no means exhaustive of the possible remedies suggested includes transition path sampling~\cite{bolh+02arpc}, 
umbrella sampling ~\cite{torri-valle77jcp,roux95cpc}, local elevation~\cite{hube+94jcamd}, Wang-Landau~\cite{wang-land01prl}, adaptive biasing force~\cite{darv-poho01jcp}, metadynamics~\cite{laio-parr02pnas,laio+05jpcb,buss+06prl,bard+11cms} and self healing umbrella sampling~\cite{mars+06jpcb}. 

We focus here on metadynamics (MetaD)~\cite{laio-parr02pnas,laio+05jpcb}.  In this approach one starts by identifying a set of appropriately chosen collective variables $s$ which are function of the microscopic variables $q$. In order to accelerate sampling,
 a bias potential is dynamically added during the simulation. The bias has the effect of helping the system to overcome large free-energy barriers so as to accelerate sampling. Asymptotically,
the negative of the bias provides an estimate of the free energy $F(s)$ associated with the collective variable $s$.
 The estimate has been demonstrated to be free of systematic errors,
if the CVs are properly chosen~\cite{buss+06prl}.

Metadynamics has been successfully used in several different contexts,
for a recent review see for instance Ref.~\cite{bard+11cms}.
 In most cases the bias is  constructed by periodically adding a repulsive Gaussian potential ~\cite{hube+94jcamd} which is a function of the $s$. This implies defining the height and width of the Gaussians. The first is related to the energy deposition rate while the latter is taken small enough  to resolve the free-energy surface features. Prescriptions on how to choose these parameters have been given and  the dependence of  the  statistical error on their choice has been discussed~\cite{laio+05jpcb}.   Still one is not guaranteed \emph{a priori} of making the best choice.

More recently,  a new flavor of metadynamics has been introduced which goes under the name  of well-tempered metadynamics
(WTMetaD).\cite{bard+08prl}
In WTMetaD the speed at which the bias is added  decreases during
the simulation. WTMetaD maintains the property that  the asymptotic bias is  related to $F(s)$ by a simple relation but, at variance with standard
metadynamics,
 the final free-energy estimate converges to a definite limit. 
 
 A useful property of WTMetaD is that the dependence of the final result on the  speed with which the bias grows is
smaller than in standard metadynamics~\cite{bard+08prl}, reducing the impact of an improper choice of this parameter.
Here we want to move further towards making the method even more efficient and robust with respect to parameters choice. A natural step in this direction is  to add the possibility of adapting the Gaussian width to the local free energy so as to speedup sampling.  For instance  a free-energy surface may present minima with  rather different curvature and a Gaussian width optimal for one  might be not appropriate for another
(see Supplementary Information, Section~\ref{sec:supplementary}).
This problem is amplified in the case of CVs that are  are highly non-linear functions of the atomic coordinates.  

In the past this need was recognized. Therefore \emph{ad hoc}  solutions for speeding up free-energy surface exploration were developed \cite{iann-laio-parr03prl}
and some form of metadynamics that uses adaptive Gaussians was implemented in  publicly available codes
\cite{CPMD}. Very recently Tribello \emph{et al}~\cite{trib-ceri-parr10pnas} also used adaptive Gaussians in a related but distinct context.
However, the effect of these choices on the free-energy reconstruction has never
been  systematically investigated.

Another common assumption in multidimensional MetaD simulations is to adopt Gaussian functions 
whose axes are aligned to the chosen CVs, i.e.~with a diagonal covariance matrix.
 This is clearly a simplification which reduces the number of input parameters, but
 an optimal choice  might  require Gaussian axes which are not aligned to the
CVs.  This was  pointed out in Ref.~\cite{iann-laio-parr03prl}
 and  successively has been used for crystal-structure prediction~\cite{mart+06natmat,mart+07prb}.

In this paper, we discuss  the possibility of performing
metadynamics simulations using multivariate
Gaussian potentials with a full covariance matrix computed on the fly.
We first introduce two possible different schemes for choosing
the  shape and width of the Gaussian potentials. One is dynamical and based on the
mean square displacement of the CVs in a predetermined time interval.  The other is geometrical and  based
on the local mean square displacement of the CVs due to a hypothetical change in the microscopic coordinates.
Subsequently we observe that the adoption 
of either of these two position-dependent covariances requires modifying the  free-energy estimator in order 
to reconstruct accurately the free-energy landscape. Furthermore, we show that this new scheme applies also when the added Gaussian potentials or some other chosen form of potentials are  coarser than the underlying free energy.
The algorithms and their efficiency are numerically tested in the standard case of alanine dipeptide in vacuum. Details of the simulations and error calculations are discussed in the Appendix.

\section{Methods}

\subsection{Metadynamics}

In MetaD, a time-dependent bias potential $V(s,t)$ 
is used to discourage the system from visiting already explored regions
in the CVs space \cite{laio-parr02pnas,bard+08prl}.
At the beginning of the simulation the bias is everywhere equal to zero.
Subsequently it is evolved according to the following expression:
\begin{equation}
\label{eq:wt-delta}
\dot{V}(s,t)=\omega e^{-V(s(t),t)/\Delta T} g(s,s(t))
\end{equation}
where $g(s,s(t))$ is a short-ranged kernel function.

Here $\Delta T$ is an energy which can be used to tune the region of free energy explored, and $\omega$ is the initial filling rate.
It has been shown in Ref.~\cite{bard+08prl} that for large times the bias is related to the free energy by
\begin{equation}
\label{eq:fes-WT}
F(s)=-\lim_{t \to \infty} \frac{\Delta T +T}{\Delta T}[V(s,t)-C(t)]~.
\end{equation}
where $C(t)$  does not depend on $s$. We notice here that even if $C(t)$
eventually diverges, its impact on the result is completely irrelevant as it
disappears in any free-energy difference calculation. The limiting behavior of $C(t)$ is discussed
in Section~\ref{sec:bias-divergence-law}.

These two expressions covers both the case of WTMetaD\cite{bard+08prl} (where $\Delta T$ is finite and 
is an additional tuning parameter) and non-WTMetaD\cite{laio-parr02pnas},
which is simply recovered in the limit
 $\Delta T \rightarrow \infty$
and corresponds to the case where the filling rate is kept constant during the simulation.
The properties of Equation~\ref{eq:wt-delta} were
extensively discussed in Ref.~\cite{bard+08prl}.
Equation~\ref{eq:fes-WT} is of practical use since it offers an estimate of the free energy 
up to an arbitrary constant
if one assumes its validity also at finite time
\begin{equation}
\label{eq:fes-old-estimator}
\tilde F(s,t)=-\frac{\Delta T +T}{\Delta T}[V(s,t)-C(t)]~.
\end{equation}

In most practical applications, the kernel $g(s,s(t))$ of Equation~\ref{eq:wt-delta}
has been replaced by
a smooth function of the CVs, typically a product of one dimensional Gaussians.
We choose here a generic multivariate Gaussian function:
\begin{multline}
\label{eq:wt-gaussian}
\dot{V}(s,t)=\omega e^{-V(s(t),t)/\Delta T} 
\times \\
\exp\left(-\frac{1}{2}\sum_{ij} [s_i-s_i(t)]\sigma_{ij}^{-2}[s_j-s_j(t)]\right) .
\end{multline}
Here, $\sigma_{ij}$ determines the shape of the Gaussian potential.
In conventional MetaD simulations, $\sigma_{ij}$ have almost always been assumed diagonal with only few exceptions
 \cite{iann-laio-parr03prl,mart+06natmat,mart+07prb,trib-ceri-parr10pnas}.
We stress once again that in all these cases MetaD was not employed to
estimate the underlying free energy, but only to speedup the exploration of configuration space and no attention has been paid to the effect of these
protocols on the free-energy reconstruction.

Passing from a diagonal $\sigma_{ij}$ to a non-diagonal one
increases the number of parameters that have to be chosen when setting up the simulation,
so that it would be convenient to give a prescription  that simplifies this choice.
In standard MetaD the diagonal elements of $\sigma_{ij}$
are obtained by computing the CV standard deviation 
during a short preliminary simulation. A simple extension is to 
consider also the correlation between different variables

\begin{equation}
\sigma_{ij}^2 \propto \langle \Delta s_i \Delta s_j \rangle~.
\label{eq:cov}
\end{equation}

This choice is completely invariant with respect to an arbitrary
linear  transformation in the CV space and reduces to a diagonal $\sigma_{ij}$ by a suitable transformation \cite{mart+06natmat}. 
In addition it allows easy mixing of CVs of different units and  nature.
However an optimal $\sigma_{ij}$ over the entire CVs domain requires that is made position dependent. Thus we shall still  use Equation~\ref{eq:cov} to estimate the Gaussian covariance but, in the same spirit of Ref.~\cite{iann-laio-parr03prl}~, we shall give a time dependent estimation of  $\langle \Delta s_i \Delta s_j \rangle$ so as to reflect the local properties of the free energy. We finally note that the use of Equation~\ref{eq:cov}  still preserves the property of  invariance relative to linear combination of the $s$ variables. However, the covariance matrix cannot be reduced to a diagonal form everywhere by means of a single linear transformation of the CVs since the matrices $\sigma_{ij}$ at different times or positions cannot be expected in general to commute.

\subsection{Dynamically-adapted Gaussian}
\label{sec:diffusion-adapted-gaussian}

To define a time-dependent adaptive covariance  at time $t$ we compute  the average value of the CVs and all the elements of the covariance matrix  from the last part of the trajectory. The center of the Gaussian is placed at the computed  average value. To select the segment of trajectory over which we perform the average we found convenient to  introduce an exponential weighting function with characteristic decay time $\tau_D$,  such that
the  Gaussians' centers  $\bar s_i(t)$  and their covariances at time $t$ are given by:
\begin{equation}
\bar s_i(t)=\frac{1}{\tau_D}\int_{0}^{t} dt^{\prime} s_i(t^{\prime})\ e^{-(t-t^{\prime})/\tau_D}
\label{eq:expavg}
\end{equation}
and 
\begin{equation}
\sigma^2_{ij}(t)= \frac{1}{\tau_D}\int_{0}^{t} dt^{\prime} [s_i(t^{\prime})-\bar s_i(t^{\prime})][s_j(t^{\prime})-\bar s_j(t^{\prime})] \ e^{-(t-t^{\prime})/\tau_D}.
\label{eq:expavg_fluct}
\end{equation}
With this choice the Gaussian location and covariance change smoothly and can be very easily evaluated.
To this effect we take the time derivative of Equation~\ref{eq:expavg} and  Equation~\ref{eq:expavg_fluct} 

\begin{equation}
\dot{\bar{s}}_i(t)=\frac{s_i(t)-\bar{s}_i(t)}{\tau_D}
\label{eq:expavg_der}
\end{equation}
\begin{equation}
\dot{\sigma}^2_{ij}(t)=\frac{[s_i(t)-\bar{s}_i(t)][s_j(t)-\bar{s}_j(t)]-\sigma^2_{ij}(t)}{\tau_D}
\label{eq:expavg_fluct_der}
\end{equation}
and then consider $\bar{s}_i(t)$ and $\sigma^2_{ij}(t)$  as additional  variables to be evolved together with 
the system dynamics. 
Integrating Equation~\ref{eq:expavg_der} and   Equation~\ref{eq:expavg_fluct_der}   with the initial conditions $\bar{s}_i(0)=s_i(0)$ and $\sigma^2_{ij}(0)=0$ the values of $\bar{s}_i(t)$ and $\sigma^2_{ij}(t)$ as defined by Equation~\ref{eq:expavg} and Equation~\ref{eq:expavg_fluct} are recovered. We name this scheme \emph{dynamically-adapted} (DA). Using partial time averages to determine $\sigma^2(t)$  is somewhat natural and  it has the  practical benefit that only one parameter $\tau_D$ determines the  whole covariance matrix.
 
Within this scheme $\tau_D$ determines the time window which is used to estimate the CV fluctuations and thus to
choose the Gaussian width. 
It is instructive to see how the latter depends on the dynamical properties of the system at least in the simplified
case of Langevin dynamics (see example in the Supplementary Information, Section~\ref{sec:supplementary}). In this case, two regimes can be identified, for short and
long value of the simulation time. At the beginning of the
simulation the dynamics is still stuck in the metastable minima. If $\tau_D$ is larger than the typical
autocorrelation time of the CV the Gaussian shape will be equal to the shape of the corresponding free-energy minimum, thus providing an optimal filling.
At the end of the simulation, when the barriers have been smoothed out by the adaptive bias, the 
dynamics is close to a free diffusion, with a
diffusion coefficient which is possibly position dependent.
Simple dimensional considerations can be used to show
that the computed $\sigma^2_{ij}$ matrix becomes proportional to the position-dependent diffusion tensor $D_{ij}(s(t))$,
which has a relevant role in describing many important phenomena.\cite{best-hummer10pnas}

\subsection{Geometry-adapted Gaussians}
\label{sec:geometry-adapted-gaussian}
We shall now discuss an alternative protocol for choosing the $\sigma$ matrices.
Our starting point will be again  Equation~\ref{eq:cov}.
For small displacements of the microscopic variables $q$ the associated change in each CV can be linearly approximated as 
\begin{equation}
\Delta s_i \approx \frac{\partial s_i}{\partial q_{\alpha}}\Delta q_{\alpha}.
\end{equation}
 Then in this approximation it can be easily seen that the CVs covariance  is linearly related to that of  the atomic  displacements $\langle\Delta q_{\alpha} \Delta q_{\beta}\rangle$. If we assume these to be  Gaussian distributed with standard deviation
  $\sigma_G$ we have
$\langle\Delta q_{\alpha} \Delta q_{\beta}\rangle=\delta_{\alpha\beta}\sigma_G^2$
leading to
\begin{equation}
\langle\Delta s_i \Delta s_j\rangle\approx
\sigma_G^2
\sum_{\alpha}
\frac{\partial s_i}{\partial q_{\alpha}}
\frac{\partial s_j}{\partial q_{\alpha}}.
\end{equation}
We use this expression to define the shape of the
Gaussian covariance via the Gram matrix: 
\begin{equation}
\label{eq:define-metric}
\sigma_{ij}^2(q)=
{\sigma^2_G}
\sum_{\alpha}
\frac{\partial s_i}{\partial q_{\alpha}}
\frac{\partial s_j}{\partial q_{\alpha}}.
\end{equation}
With this choice,
$\sigma^2(q)$ depends explicitly on the microscopic variables $q$.
We shall refer to this case  as \emph{geometry-adapted} (GA). As in the DA case only one parameter suffices to determine the whole covariance matrix.
  We note that in MetaD the Gaussian spread plays a  role similar to that of the histogram bin size for other methods and determines which configurations are considered as equivalent.
Thus here  we are assuming as equivalent all the microscopic configurations whose a root square distance of the atoms involved in the CVs is within $\sigma_G$ making possible for the choice of  $\sigma_G$ to be guided  by physical considerations.
For instance, if one were to choose a different set of CVs
still the choice of sigma would be determined by the typical
atomic displacements.

Similarly to the DA scheme, the GA one is invariant with respect to linear transformations of the CVs.

\section{Free-energy estimation}

One of the most interesting features of metadynamics
is the link it establishes between the free-energy surface and the bias both in its classical and well-tempered versions. Having changed the protocol  of Gaussians deposition it is crucial to establish such a link in the case of Gaussians of variable covariance.

\subsection{Small-width limit}
\label{sec:small-width-limit}
We  consider first the case in which the Gaussians have a size  smaller than the relevant features in the free-energy surface, as normally done in MetaD. This was the assumption of  Ref.~\cite{bard+08prl} where one went as far as to assimilate the Gaussians to $\delta$ functions.

Within our schemes this limit can be achieved by choosing $\tau_D$ or $\sigma_G$ small enough. In this regime, for the DA scheme, one can also neglect the difference between the actual position $s$ and the center of the deposited Gaussian $\bar s$.
Therefore the only adjustment necessary is to take into account the fact that the change in covariance induces a change in Gaussians' volume.
Thus in Equation~\ref{eq:wt-delta}  we replace the kernel $g(s,s(t))$
with a Dirac delta with the same normalization, $\sqrt{(2\pi)^d}\det\sigma\delta(s-s(t))$,
 where $\sigma$ is either a function of the trajectory in CVs' space
(DA scheme) or of the microscopic coordinates at a specific time (GA scheme)
and $d$ is the number of CVs: 
\begin{equation}
\label{eq:wt-general}
\dot{V}(s,t)= \omega e^{-V(s(t),t)/\Delta T}\sqrt{(2\pi)^d}\det\sigma\delta(s-s(t)).
\end{equation}
As in Ref.~\cite{bard+08prl}, we notice that for large times
the probability distribution becomes
$P(s,t)ds\propto\exp\left(
-\frac{F(s)+V(s,t)}{T}
\right)$ and one has
\begin{multline}
\label{eq:wt-average}
\dot{V}(s,t)=\omega e^{-V(s,t)/\Delta T}\sqrt{(2\pi)^d} \langle \det\sigma \rangle_{s}P(s,t) 
\propto \\ e^{-V(s,t)/\Delta T}\ \langle \det\sigma \rangle_{s}
\ e^{-[F(s)+V(s,t)]/T}
\end{multline}
where the average $\langle \det\sigma \rangle_{s}$ is taken in the canonical
ensemble at a fixed value of the Gaussian center.
Now, setting $\dot{V}=C(t)$ in the last equation, where $C(t)$ is constant with respect to
the CVs, we find that
the asymptotic solution for $V(s,t)$ is
\begin{equation}
\lim_{t\rightarrow\infty}
V(s,t)
=
-\frac{\Delta T}{\Delta T+T}\left[
F(s)-T\ln\langle \det\sigma \rangle_{s}
\right]
+C(t)~.
\label{eq:bias-new}
\end{equation}
This means that the bias is no longer proportional to $F(s)$  as in Equation~\ref{eq:fes-WT} but rather to
\begin{equation}
G(s)=F(s)-T\ln\langle \det\sigma \rangle_{s}.
\label{eq:gauge-invariant}
\end{equation}
In the case of GA scheme of Section~\ref{sec:geometry-adapted-gaussian},
$G(s)$ turns out to be the gauge invariant free-energy
discussed in
Refs.~\cite{e-vand04book,vand-tal05jcp,hart-schut07pdnp}
which is not changed by an
arbitrary non-linear monotonic transformation of the CVs.
Instead in the DA case of Section~\ref{sec:diffusion-adapted-gaussian} when $\tau_D$ is chosen so that $\sigma$ is proportional to the diffusion matrix, our bias potential is closely connected with that used in flux-tempered metadynamics  \cite{sing+11jsp}  which is also gauge invariant and has been designed to optimize the round trip time.

While gauge invariance is esthetically pleasing it does not bring particular advantages since for physical applications it is $F(s)$ that is needed.
This can be simply obtained  by rewriting  Equation~\ref{eq:bias-new} as
\begin{equation}
\label{eq:fes-geometric}
F(s)=-\lim_{t\rightarrow\infty}\frac{\Delta T +T}{\Delta T}[V(s,t)-C(t)]+T\ln\langle \det\sigma \rangle_{s}
\end{equation}
which is the  appropriate generalization  of Equation~\ref{eq:fes-WT} to the case of small adaptive Gaussians.

\subsection{Free energies from reweighting}
\label{free_from_reweight}

As discussed in the previous sections, for non-standard biasing protocols
 the relationship between the 
asymptotic bias and the underlying free-energy landscape is not known \emph{a priori}.
Only in the small width limit it is possible to estimate  explicitly the correction of Equation~\ref{eq:fes-geometric}.
However it would be nice to have an estimator that is valid for large Gaussians.
 This is provided by the  relation:
\begin{equation}
\label{eq:fes-estimator}
\tilde{F}_N(s,t)=-T\ln N(s,t)-V(s,t)+T\log \int ds' N(s',t)
\end{equation}
where $N(s,t)$ is the accumulated histogram of the variable $s$ up to time $t$.
If the bias is time-independent this relation is strictly true and is normally used in umbrella sampling. 
 In this present context it is valid only for large times 
 when $V(s,t)$ has converged. More precisely it is only necessary that the rate of bias deposition goes to zero sufficiently rapidly. Therefore it will be valid whether the Gaussian width is small or large and whether the width is determined by diffusion or geometry.

This relation
was also used in deriving
a free-energy estimator for WTMetaD~\cite{bard+08prl} where this expression was manipulated
in a manner similar to Section~\ref{sec:small-width-limit}
 leading to
the  estimator 
in Equation~\ref{eq:fes-old-estimator}.
However, $\tilde{F}_N(s,t)$ is more generally valid  and gives
a correct estimate of free energy even when adaptive Gaussians
are used.
In the following we will show how this new estimator can be of use in a practical case.

The case of non-WT metadynamics (i.e. for $\Delta T\rightarrow\infty$)  where the bias 
 oscillates in the long time limit needs separate consideration and will be discussed elsewhere. 

\section{Examples}

\subsection{Narrow and wide Gaussians }\label{sec:small_vs_large}

\begin{figure}
\begin{center}
\includegraphics[scale=0.25]{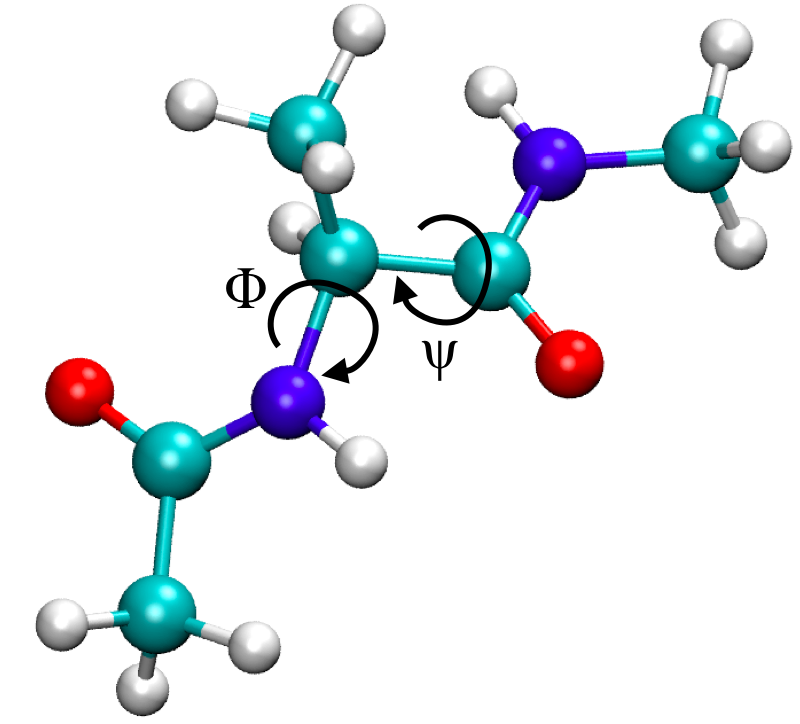}
\caption{Molecular sketch of alanine dipeptide with the Ramachandran dihedral angles $\Phi$ and $\Psi$. This graphics was produced with VMD\cite{VMD}.}
\label{fig:diala}
\end{center}
\end{figure}
Before discussing the performance of the adaptive Gaussians we shall first
validate the estimator of Equation~\ref{eq:fes-estimator} when using standard
WTMetaD and compare two very different choices of Gaussian width.
In this way we show that it is advantageous to use
 Equation~\ref{eq:fes-estimator} even when using Gaussians with fixed covariance.

To this effect we performed a number of WTMetaD simulations of
alanine dipeptide in vacuum (see Figure~\ref{fig:diala}) using as collective variables  the two Ramachandran \cite{ramachandran} dihedral angles $\Phi$ and $\Psi$.  The initial Gaussian height was chosen to be 0.287 $\rm kcal~mol^{-1}$  and Gaussian potentials were deposited every 120 fs, corresponding to an energy deposition rate of 2.39 $\rm 10^{-3}\ kcal\ mol^{-1}\ fs^{-1}$. The $\Delta T$ parameter for WTMetaD was set to 1200 K. Other technical details can be found in Section~\ref{sec:simulation-details}.

We first performed a long calculation to obtain a reference free-energy landscape. This was done in two steps: we performed a
5 ns long WTMetaD run with a Gaussian width $\sigma$= 0.35 rad for both $\rm \Phi$ and $\rm \Psi$ and the bias thus accumulated was then kept constant in 
a long (1$\mu$s)  biased simulation.  Then, by using Equation~\ref{eq:fes-estimator} which 
is exact for a static bias, we obtained the reference free energy.
The error of the reference landscape was estimated by comparing the free energy derived from the histogram  from the first and 
the second half of the simulation and was obtained using the procedure
reported in Section~\ref{sec:error-estimate}. The error is approximately 0.01 kcal mol$^{-1}$, which is negligible
as compared to the  error of more than 0.05 kcal mol$^{-1}$ made in the test runs described below. 
We did not bin the $s$ to evaluate the histogram $N(s,t)$  in Equation~\ref{eq:fes-estimator} but instead  we used  Gaussian functions
of width 0.01 rad. This led to a smooth $N(s,t)$ and a smooth free energy as shown in Figure~\ref{fig:panel_rama}A.

We then  performed two different sets of simulations using the same $\sigma$ for both dihedral angles. In one case  $\sigma$ 
was smaller than the free-energy features ($\sigma$= 0.35 rad), in the other larger ($\sigma$= 0.7 rad). 
Each set of simulations consisted of 100 WTMetaD runs of 20 ns each. The error relative to the reference free energy  committed using the standard estimator  of Equation~\ref{eq:fes-old-estimator} and the new one in Equation~\ref{eq:fes-estimator} was then compared.

\begin{figure}
\begin{center}
\includegraphics[scale=0.25]{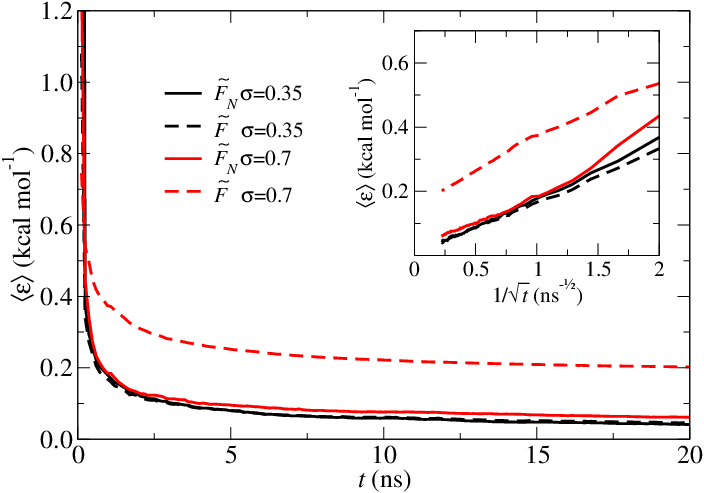}
\caption{The estimate of the average error as a function of simulation time for alanine dipeptide in vacuum. The average error was calculated through a set of 100 runs and averaged every 120 ps. The effect of adopting  $\tilde F_N$ of Equation~\ref{eq:fes-estimator} for two different choices of $\sigma$ is compared to the choice of  $\tilde F$ of  Equation~\ref{eq:fes-old-estimator}. It is possible to appreciate that Equation~\ref{eq:fes-estimator}  always delivers results that are comparable to the  Equation~\ref{eq:fes-old-estimator} or better, when large Gaussians are adopted. In the inset it is shown the time dependence of the error that decreases linearly with $1/\sqrt{t}$ predicting an error that goes to zero for infinite time. On the contrary, for large Gaussians, the standard estimate is not able to resolve the features of the free energy resulting in a significant residual error even for infinite time.}
\label{fig:small_vs_big}
\end{center}
\end{figure}

From Figure~\ref{fig:small_vs_big} it can be seen that  in all cases the new estimator 
produces consistent results and the expected
 asymptotic  $1/\sqrt{t} $  behavior. The old one works well for small Gaussians but in the case of the
large ones is affected by a systematic error, due to the fact that it is not able to
resolve the smaller features of the free energy landscape.

\subsection{Dynamically-adapted Gaussians}

Having assessed  the usefulness of Equation~\ref{eq:fes-estimator} we turn to the evaluation of the performance  
of  Gaussians of variable covariance using the DA scheme. 
This  requires defining the parameter $\tau_D$ that is a measure of the 
time required to sample two bins  that we consider as different. 
 In order to
choose $\tau_D$ in a manner which allows a fair comparison of constant and variable Gaussians covariance
we performed a preliminary 1 ns of standard WTMetaD with Gaussians of a constant  $\sigma$=0.35 rad width.
We then tested different $\tau_D$ values and we choose the one that  was able to fill on average the same volume deposited in the standard run. In this way we ensured that the system was subject to a comparable filling rate. This gave the value  $\tau_D$= 300 fs.
In analogy to the previous protocol, we performed 100 runs of 20 ns each and calculated the average error made in the region of free-energy surface which is within lowest 5 $\rm kcal\ mol^{-1}$ from  the minimum of the reference surface.

The results are displayed in Figure~\ref{fig:err_phi_psi}, where
the failure of  
the old estimator of Equation~\ref{eq:fes-old-estimator} is evident.
Instead the new estimator gives correct results even when the covariance of the Gaussians is let to vary.

\subsection{Geometry-adapted Gaussians: Ramachandran plot}

\begin{figure}
\begin{center}
\includegraphics[scale=0.25,angle=0]{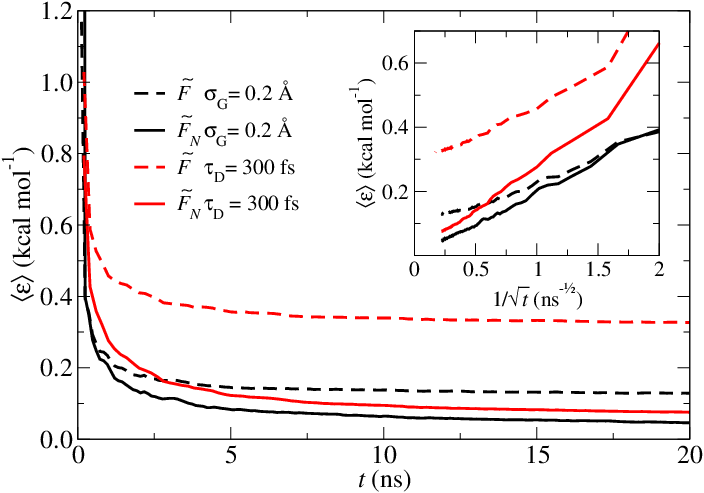}
\caption{Calculation of the average error for both GA (denoted with $\sigma_G$ ) and DA (denoted with $\tau_D$) schemes for alanine dipeptide as a function of time. The average error was calculated through a set of 100 runs and averaged every 120 ps. Two free-energy estimates are employed: the one of Equation~\ref{eq:fes-estimator} denoted with $\tilde F_N$ and the one of Equation~\ref{eq:fes-old-estimator} denoted with $\tilde F$. The performance of Equation~\ref{eq:fes-estimator} is always superior to that of  Equation~\ref{eq:fes-old-estimator} which is dramatically incorrect when DA or GA are employed. In the inset it can be appreciated that the error scales linearly with $1/\sqrt{t}$ as expected in WTMetaD. The use of Equation~\ref{eq:fes-estimator} leads to the correct asymptotic behavior in contrast to Equation~\ref{eq:fes-old-estimator}.}
\label{fig:err_phi_psi}
\end{center}
\end{figure}

We now turn to a test of the GA scheme for the same system using an identical protocol. The choice of a comparable $\sigma_G$ was performed as before imposing the condition of the same volume filling rate with respect to a preliminary standard WTmetaD. This led to  a value  $\sigma_G$ =0.2\ \AA.
The results are again displayed in Figure~\ref{fig:err_phi_psi}.
The free energy calculated through $\tilde F$ of Equation~\ref{eq:fes-old-estimator} is expected to converge to the gauge invariant free-energy landscape,
which is different from the standard one. Thus, even for long times, there is a systematic residual error.
On the other hand, the new estimator  does give the correct result,
with a convergence rate which is very similar to that obtained with the standard calculation reported in Figure~\ref{fig:small_vs_big}.

\begin{figure*}
\begin{center}
\includegraphics[scale=0.36]{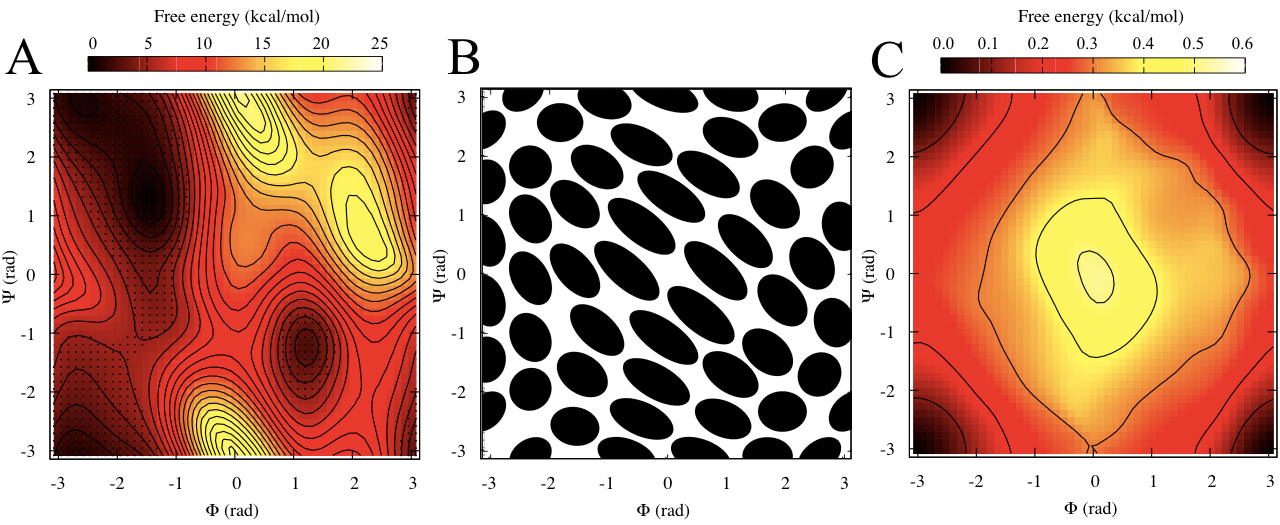}
\caption{Panel A: free-energy landscape on the two Ramachandran dihedral angles $\Phi$ and $\Psi$. The dotted region represents the portion used for  error calculations which lies within 5 $\rm kcal~mol^{-1}$ from the lowest free-energy point. 
Panel B: representative sketch of the shape of the Gaussians produced on the Ramachandran plot for alanine dipeptide obtained by using the GA scheme of Section~\ref{sec:geometry-adapted-gaussian}. The size of each ellipse is scaled so to reflect the average size of the Gaussian potentials placed in each specific point.
Panel C: free-energy contribution coming from the change in volume of the Gaussians induced by the adoption of the GA scheme.
\label{fig:panel_rama}
}
\end{center}
\end{figure*}

In Figure~\ref{fig:panel_rama}B,  we give a pictorial view of the shape of the Gaussian potentials obtained by  using Equation~\ref{eq:define-metric}. We also report in  Figure~\ref{fig:panel_rama}C the correction due to the change in volume of the Gaussians, which measures the error made by using  the old estimator. It can be seen that in this case the maximum error is as small as ~0.6 $\rm kcal\ mol^{-1}$. However this is by no means generally the case as we shall see in the next section.

\subsection{Geometry-adapted Gaussians: double MSD}

We examine now the role that a different choice of collective variables might have on the error made by using the old estimator. While for the angle $\Phi$ and $\Psi$ the error was small we expect it to be larger when dealing with CVs that are strongly nonlinear functions of the atomic coordinates. One such example  are  for instance CVs that measure the mean square deviation (MSD) from a reference structure or coordination functions that have sigmoidal dependence on the distance of two atoms or groups of atoms. 

\begin{figure*}
\begin{center}
\includegraphics[scale=0.36]{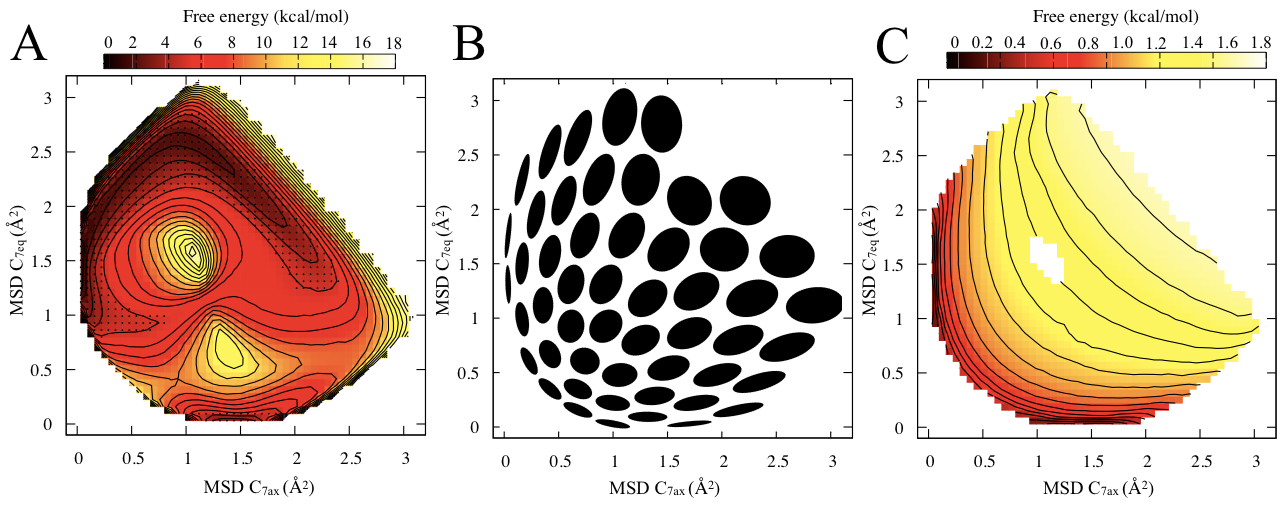}
\caption{
Panel A:  free-energy landscape on the MSD from the two metastable basins $\rm C_{7eq}$ and $\rm C_{7ax}$. The dotted region represents the portion used for  error calculations which lies within 5 $\rm kcal~mol^{-1}$ from the lowest free energy point.
Panel B: representative sketch of the shape of the Gaussians produced by using two MSD as CVs. The Gaussian widths were obtained by the GA scheme of Section~\ref{sec:geometry-adapted-gaussian}. The size of each ellipse is scaled so to reflect the average size of the Gaussian potentials placed in that specific point.
Panel C: free-energy contribution from the change in volume of the Gaussians induced by the adoption of the GA scheme. The blank region in the center is due to the lack of sampling for this simulation time. }
\label{fig:panel_msd}
\end{center}
\end{figure*}

Thus we studied once again alanine dipeptide in vacuum and used as CVs  the MSD from the two conformers $\rm C_{7eq}$ and $\rm C_{7ax}$ rather than the two torsional angles. The MSDs were calculated through optimal alignment by using Kearsley's algorithm \cite{kearsley} and only the heavy atoms were considered in the metrics.
In order to obtain reference values as accurate as possible rather than reweighting the configurations previously generated,  we repeated the simulation with the same protocol as before. The result of this run which also lasted  1$\mu$s  is shown in Figure~\ref{fig:panel_msd}A where it can be seen that in these new variables the free energy surface has two minima one very narrow and the other wider.

\begin{figure}
\begin{center}
\includegraphics[scale=0.25,angle=0]{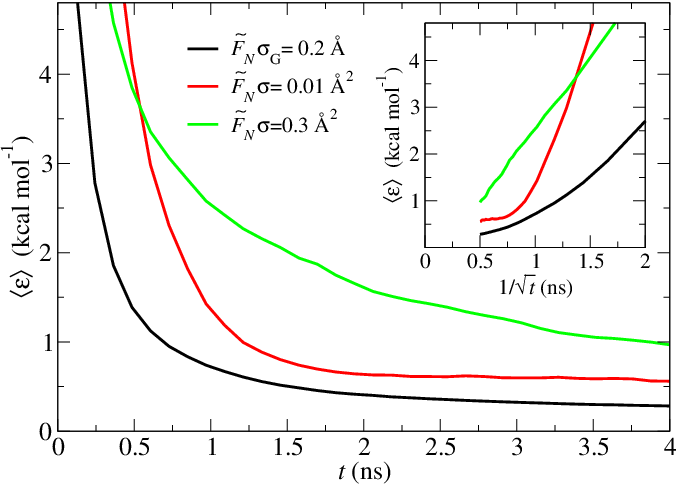}
\caption{WTMetaD for the double MSD case for alanine dipeptide in vacuum. For each choice of $\sigma$, 100 runs of 4 ns each were performed
and the average error respect to the reference free energy was calculated as function of time. Two different choices of $\sigma$ are made and compared with the result obtained using  the GA scheme with Gaussian width $\sigma_G$. The GA scheme is always superior and converges faster to the correct value.} 
\label{fig:big_small_rmsd}
\end{center}
\end{figure}

By comparing Figure~\ref{fig:panel_msd}B and Figure~\ref{fig:panel_rama}B it is evident that the double MSD space induces a much larger change in shapes and volumes of the deposited Gaussian potentials. Correspondingly the error made using the old estimator of Equation~\ref{eq:fes-old-estimator} becomes larger (see Figure~\ref{fig:panel_msd}C).
The benefit of using variable Gaussians becomes very apparent since we do not have to choose  small Gaussians to resolve the  narrow minima paying the price of a  slow convergence,
nor do we have  to use larger Gaussians and sacrifice accuracy.
This is exemplified in Figure~\ref{fig:big_small_rmsd} where we show the typical convergence behavior obtained by using the GA scheme of Equation~\ref{eq:define-metric} and compare it
to fixed-$\sigma$ runs, where $\sigma$ was chosen to be appropriate either to the narrow minimum ($\sigma$=0.01 \AA $\rm^2$) or to the larger one ($\sigma$=0.3 \AA $\rm^2$).
Using the new estimator all three calculation appear to converge to the same limit, but with different rates.
For narrow Gaussians ($\sigma$=0.01 \AA$\rm^2$) the calculation converges to the right limit albeit rather slowly. Similarly for very large Gaussians ($\sigma$=0.3 \AA$\rm^2$) the convergence was very slow for reasons similar to those discussed earlier
(see Section~\ref{sec:small_vs_large}).
When the GA scheme of Section~\ref{sec:geometry-adapted-gaussian} is chosen the asymptotic limit is reached faster and with a much smaller error. 
 
\section{Discussion and conclusions}

In summary, we have investigated in detail the possibility
of performing metadynamics simulations where the repulsive Gaussian potentials
have a width chosen on the fly and are not aligned with respect to the CVs.
We have shown that using adaptive widths can lead to artifacts in the estimation
of the free-energy landscape, which can be recovered by using a suitable
estimator. Moreover, we have discussed two independent recipes to
adapt the Gaussian shape, one based on the time evolution of the collective variables and another based on the intrinsic metrics of the microscopic
coordinates.
Both methods
can be implemented using the same ingredients as standard metadynamics
calculations, namely collective-variable values and first
derivatives with respect to atomic positions.
They are effective and
have been shown to produce unbiased results
on a standard benchmark.
 Additionally they can remarkably improve the
filling speed especially when collective variables which are highly non-linear functions
for the atomic positions are used. This is particularly helpful for MSD and MSD-based variables
\cite{bran+07jcp}
or contact map-based variables which are widely employed for systems of biological and condensed matter interest. For such variables, the usual choice of the Gaussian width fixed once and for all to the value of  the CVs fluctuation measured in an initial unbiased run may in fact turn to be suboptimal whenever the system is far from its initial configuration. In this respect the gradient-adapted choice is of great help.
These methods simplify the choice of the input parameters for a metadynamics simulation,
significantly reducing the time which is usually spent in trial and errors.
Both approaches can be used with CVs which are bounded, either by their intrinsic definition or by artificially added potentials.
In the latter case, the version based on time evolution of the CVs might be more effective.
Finally, they both can be straightforwardly  combined with versions of MetaD
which are based on multiple replicas~\cite{rait+06jpcb,buss+06jacs,pian-laio07jpcb}.

\section*{Acknowledgement}
The work has been performed under the HPC-EUROPA2 project (project number: 228398) with the support of the European Commission - Capacities Area - Research Infrastructures.
G.B.~acknowledges MIUR grant ``FIRB - Futuro in Ricerca'' no. RBFR102PY5 for funding.

\appendix

\section{Bias divergence law}
\label{sec:bias-divergence-law}

The relation between the absolute free energy and the bias [Equation~\ref{eq:fes-WT} and Equation~\ref{eq:fes-geometric}] contains a
  shift $C(t)$ which does not depend on $s$ but grows with the time.
As already noted, this constant is irrelevant in the calculation of free-energy differences.
However, it is instructive to compute explicitly its behavior in the long time limit. In the following
we shall do it in the small-width approximation,
which is usually applied in the analysis of WTMetaD simulations.
According to Equation~\ref{eq:wt-average} the time derivative of the bias is
\begin{equation}
\label{eq:v-dot}
\dot{V}(s,t)=\frac{\omega e^{-V(s,t)/\Delta T}\ \sqrt{(2\pi)^d} \langle \det\sigma\rangle_s e^{-\frac{F(s)+V(s,t)}{T}}}{
  \int ds' e^{-\frac{F(s')+V(s',t)}{T}} } ~.
\end{equation}
As in the long time limit the bias grows uniformly in $s$, its time derivative is equal to
the one of $C(t)$.
By combining  Equation~\ref{eq:v-dot} and Equation~\ref{eq:bias-new} one obtains
\begin{equation}
\label{eq:c-dot}
\dot{C}(t)=\frac{\omega \sqrt{(2\pi)^d} e^{-\frac{C(t)}{\Delta T}}}{
\int ds' \langle \det\sigma\rangle_{s'}^{-\frac{\Delta T}{T+\Delta T}}
   e^{-\frac{F(s')}{T+\Delta T}}}\propto e^{-\frac{C(t)}{\Delta T}} ~.
\end{equation}
By solving the differential equation Equation~\ref{eq:c-dot}, it can be seen that $C(t)$ diverges logarithmically
with time as
\begin{equation}
\lim_{t\rightarrow\infty} \frac {C(t)}{\Delta T \log t} = 1 ~.
\end{equation}
We underline that this is just the limiting behavior of $C(t)$. If its actual value is needed,
for instance to align estimates of the free energy made at different times,
the procedure illustrated in Ref.~\cite{bono+09jcc} could be used to estimate it. Note however that in  Ref.~\cite{bono+09jcc} a different sign convention is used.

\section{Simulation details}
\label{sec:simulation-details}

In all the simulations shown we employed alanine dipeptide (ACE-ALA-NME) in vacuum as
a model system, with molecular interactions described by the CHARMM27\cite{charmm} force field.
This system (see Figure~\ref{fig:diala})
is a widely known benchmark for free-energy calculations \cite{mara-fish-vand06jcp,dialanine1,dialanine2,dialanine3,dialanine4,dialanine5}.
Indeed, it displays two main basins, namely $\rm C_{7eq}$ and $\rm C_{7ax}$ separated by a sizable barrier of several $k_BT$ at 300K.
A timestep of 2 fs was employed and all
the covalent bonds involving an hydrogen atom were constrained to the equilibrium distance by means of SHAKE\cite{shake} algorithm.   
A Langevin thermostat was used with a temperature of 300K and a damping factors of 5 $\rm ps^{-1}$. NAMD 2.8\cite{namd} molecular dynamics code was used and modified to add the adaptive shape WTMetaD capability.

\section{Error calculation}
\label{sec:error-estimate}

Here we describe the procedure adopted to evaluate the error between two free-energy landscapes for the numerical examples reported in the main text.

We first note that WTMetaD for finite $\Delta T$ produces a more accurate histogram in low free-energy regions. Therefore we selected a reference region in CVs space defined by those points  lying within a value of $v$ free-energy units respect to the minimum of one of the free-energy surfaces, here termed reference free energy $F_r(s)$. In all the calculations performed the value of  $v$ was set to be 5 $\rm kcal~mol^{-1}$.

In particular, being $F(s)$  the free-energy surface whose error is required and  $s$ a point of the collective variables space, the error between the two surfaces is defined as:
\begin{eqnarray}
\epsilon&=& 
\sqrt{ \frac{\int_{S}  [ \bar F_r(s) - \bar F(s)  ]^2 \ \theta(v- F_r(s) )\ d s }{ \int_{S}\ \theta(v-  F_r(s))  d s }  }.
\end{eqnarray}
Here $S$ is the multidimensional space in which the calculation is performed, being $\Phi$ and $\Psi$ for the Ramachandran plot, and $\theta$ is a Heaviside step function. 
This equation amounts in calculating the average squared root difference between the two free energies $\bar F_r(s)$ and $\bar F(s)$ in the CVs space defined within $v$ $\rm kcal~mol^{-1}$ from the minimum in the reference free energy $\bar F_r(s)$.
The $\bar F_r(s)$ is related to $F_r(s)$ by a rigid shift of the free-energy surface respect to the average value in the reference region:
\begin{eqnarray}
\bar F_r(s)&=&F_r(s)-\frac{ \int_{S} \ F_r(s) \theta(v-  F_r(s)) \ d s }{  \int_{S}\theta(v-  F_r(s)) d s  } 
\end{eqnarray}
and a similar relation holds for the other free energy $F(s)$ where, for consistency, the reference region is again defined on the reference free energy
\begin{eqnarray}
\bar F(s)&=&F(s)-\frac{ \int_{S} \ F(s) \theta(v-  F_r(s)) \ d s }{  \int_{ S}\theta(v-  F_r(s)) d s  }.
\end{eqnarray}

\section{Supplementary Information: a simple illustrative toy model potential}
\label{sec:supplementary}

\begin{figure}[htbp]
\begin{center}
\includegraphics[width=\columnwidth]{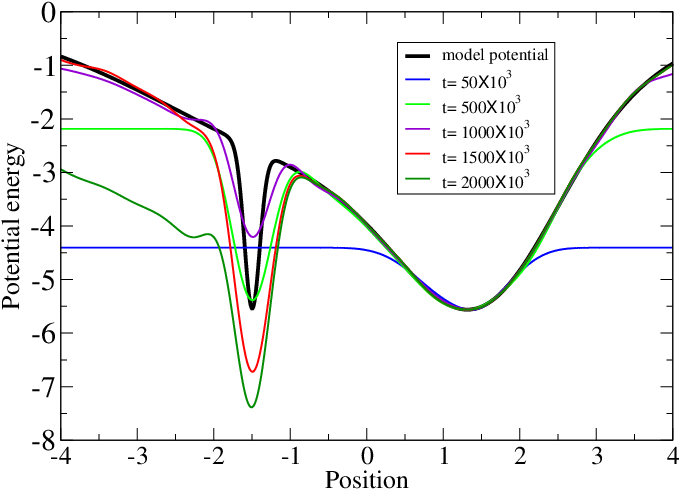}
\caption{WTMetad run where $\sigma_{WT}$=0.26 and stays constant along time. In solid black the energy profile of the model potential is reported. In color different estimates of the free-energy landscape along the WTMetaD run are reported. All the free-energy profiles are aligned to the free-energy value of the rightmost minimum in the model potential. }
\label{large}
\end{center}
\end{figure}

We want here to show how the need of using adaptive Gaussians
arises from a simple one-dimensional Langevin model.
Consider an energy landscape composed by the sum of three Gaussian potentials ($x_0$=-1.5, $\sigma_0$=0.1, $w_0$=-3, $x_1$=1.5, $\sigma_1$=1.0, $w_1$=-3, $x_2$=0, $\sigma_2$=2.5, $w_2$=-3), namely:
\begin{equation}
U(x)=\sum_{i=0}^{2} w_i\ \exp{\left( -\frac{(x-x_i)^2}{2 \sigma_{i}^2} \right)}
\end{equation}
and a particle of unitary mass that moves in this potential with a temperature of 0.2 energy units. The timestep was set to 0.0025 time units and the friction was set to 0.8 inverse time units.
The particle is initially placed in the rightmost minimum (around 1.5) and the evolution is observed for $\rm 10^6$ steps of Langevin dynamics.
After this, the standard deviation of the position is calculated and used as width ($\sigma_{WT}$=0.26) for a Well-Tempered Metadynamics (WTMetaD) run during $\rm 2\times10^6$ steps of dynamics. Additional parameters for the WTMetaD are a $\Delta T$=2.8 and a energy deposition rate of $\rm 5\times10^{-5}$ energy units/step.  
In Figure~\ref{large} the result of this WTMetaD is shown along with the model potential (solid black line). A set of estimates of the underlying free-energy landscape (identical to the potential energy in this one-dimensional case) along the time of WTMetaD run are reported and aligned to the value of potential energy of the rightmost minimum. 
It is evident that the $\sigma_{WT}$ which was suitable for the larger minimum is not suitable for the minimum on the left side thus seriously affecting both the free-energy difference between the minima as well as the barrier estimate.

\begin{figure}[htbp]
\begin{center}
\includegraphics[width=\columnwidth]{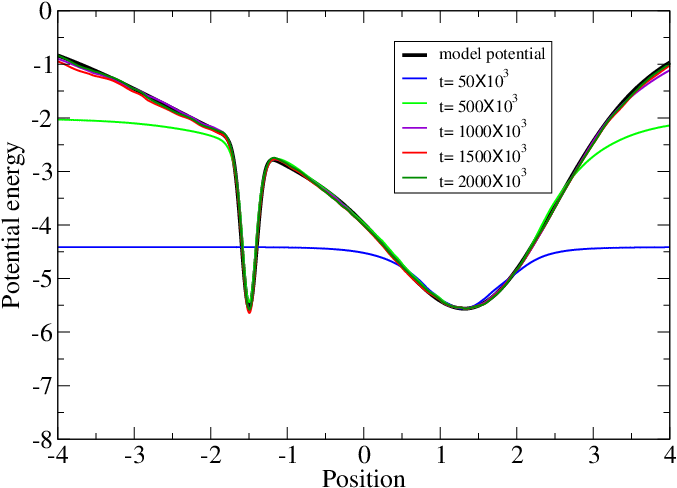}
\caption{WTMetad run where $\sigma_{WT}$ is adapted to the displacement along the last $\tau_D$=250 steps of Langevin dynamics. In solid black 
the energy profile of the model potential is reported. In color different estimates of the free-energy landscape along the WTMetaD run are reported. All the free-energy profiles are aligned to the free-energy value of the rightmost minimum in the model potential.}
\label{taud}
\end{center}
\end{figure}

On the contrary, in Figure~\ref{taud} we report results obtained by setting the $\sigma_{WT}$ equal to 
the displacement of the system in the last 250 steps of the Langevin dynamics by means of our dynamical adapted algorithm
(see main text).
Here the narrow features of the free-energy are adequately resolved and the free-energy profiles and barrier result improved and present a faster  convergence.

\providecommand*\mcitethebibliography{\thebibliography}
\csname @ifundefined\endcsname{endmcitethebibliography}
  {\let\endmcitethebibliography\endthebibliography}{}


\begin{mcitethebibliography}{40}
\providecommand*\natexlab[1]{#1}
\providecommand*\mciteSetBstSublistMode[1]{}
\providecommand*\mciteSetBstMaxWidthForm[2]{}
\providecommand*\mciteBstWouldAddEndPuncttrue
  {\def\EndOfBibitem{\unskip.}}
\providecommand*\mciteBstWouldAddEndPunctfalse
  {\let\EndOfBibitem\relax}
\providecommand*\mciteSetBstMidEndSepPunct[3]{}
\providecommand*\mciteSetBstSublistLabelBeginEnd[3]{}
\providecommand*\EndOfBibitem{}
\mciteSetBstSublistMode{f}
\mciteSetBstMaxWidthForm{subitem}{(\alph{mcitesubitemcount})}
\mciteSetBstSublistLabelBeginEnd
  {\mcitemaxwidthsubitemform\space}
  {\relax}
  {\relax}

\bibitem[Bolhuis et~al.(2002)Bolhuis, Chandler, Dellago, and
  Geissler]{bolh+02arpc}
Bolhuis,~P.; Chandler,~D.; Dellago,~C.; Geissler,~P. \emph{Annu. Rev. Phys.
  Chem.} \textbf{2002}, \emph{53}, 291--318\relax
\mciteBstWouldAddEndPuncttrue
\mciteSetBstMidEndSepPunct{\mcitedefaultmidpunct}
{\mcitedefaultendpunct}{\mcitedefaultseppunct}\relax
\EndOfBibitem
\bibitem[Torrie and Valleau(1977)Torrie, and Valleau]{torri-valle77jcp}
Torrie,~G.~M.; Valleau,~J.~P. \emph{J.~Comput.~Phys.} \textbf{1977}, \emph{23},
  187--199\relax
\mciteBstWouldAddEndPuncttrue
\mciteSetBstMidEndSepPunct{\mcitedefaultmidpunct}
{\mcitedefaultendpunct}{\mcitedefaultseppunct}\relax
\EndOfBibitem
\bibitem[Roux(1995)]{roux95cpc}
Roux,~B. \emph{Comp. Phys. Comm.} \textbf{1995}, \emph{91}, 275--282\relax
\mciteBstWouldAddEndPuncttrue
\mciteSetBstMidEndSepPunct{\mcitedefaultmidpunct}
{\mcitedefaultendpunct}{\mcitedefaultseppunct}\relax
\EndOfBibitem
\bibitem[Huber et~al.(1994)Huber, Torda, and van Gunsteren]{hube+94jcamd}
Huber,~T.; Torda,~A.~E.; van Gunsteren,~W.~F. \emph{J. Comput.-Aided Mol. Des.}
  \textbf{1994}, \emph{8}, 695--708\relax
\mciteBstWouldAddEndPuncttrue
\mciteSetBstMidEndSepPunct{\mcitedefaultmidpunct}
{\mcitedefaultendpunct}{\mcitedefaultseppunct}\relax
\EndOfBibitem
\bibitem[Wang and Landau(2001)Wang, and Landau]{wang-land01prl}
Wang,~F.; Landau,~D.~P. \emph{Phys.~Rev.~Lett.} \textbf{2001}, \emph{86},
  2050--2053\relax
\mciteBstWouldAddEndPuncttrue
\mciteSetBstMidEndSepPunct{\mcitedefaultmidpunct}
{\mcitedefaultendpunct}{\mcitedefaultseppunct}\relax
\EndOfBibitem
\bibitem[Darve and Pohorille(2001)Darve, and Pohorille]{darv-poho01jcp}
Darve,~E.; Pohorille,~A. \emph{J.~Chem.~Phys.} \textbf{2001}, \emph{115},
  9169--9183\relax
\mciteBstWouldAddEndPuncttrue
\mciteSetBstMidEndSepPunct{\mcitedefaultmidpunct}
{\mcitedefaultendpunct}{\mcitedefaultseppunct}\relax
\EndOfBibitem
\bibitem[Laio and Parrinello(2002)Laio, and Parrinello]{laio-parr02pnas}
Laio,~A.; Parrinello,~M. \emph{Proc.~Natl.~Acad.~Sci.~U.S.A.} \textbf{2002},
  \emph{99}, 12562--12566\relax
\mciteBstWouldAddEndPuncttrue
\mciteSetBstMidEndSepPunct{\mcitedefaultmidpunct}
{\mcitedefaultendpunct}{\mcitedefaultseppunct}\relax
\EndOfBibitem
\bibitem[Laio et~al.(2005)Laio, Rodriguez-Fortea, Gervasio, Ceccarelli, and
  Parrinello]{laio+05jpcb}
Laio,~A.; Rodriguez-Fortea,~A.; Gervasio,~F.~L.; Ceccarelli,~M.; Parrinello,~M.
  \emph{J.~Phys.~Chem.~B} \textbf{2005}, \emph{109}, 6714--6721\relax
\mciteBstWouldAddEndPuncttrue
\mciteSetBstMidEndSepPunct{\mcitedefaultmidpunct}
{\mcitedefaultendpunct}{\mcitedefaultseppunct}\relax
\EndOfBibitem
\bibitem[Bussi et~al.(2006)Bussi, Laio, and Parrinello]{buss+06prl}
Bussi,~G.; Laio,~A.; Parrinello,~M. \emph{Phys.~Rev.~Lett.} \textbf{2006},
  \emph{96}, 090601\relax
\mciteBstWouldAddEndPuncttrue
\mciteSetBstMidEndSepPunct{\mcitedefaultmidpunct}
{\mcitedefaultendpunct}{\mcitedefaultseppunct}\relax
\EndOfBibitem
\bibitem[Barducci et~al.(2011)Barducci, Bonomi, and Parrinello]{bard+11cms}
Barducci,~A.; Bonomi,~M.; Parrinello,~M. \emph{Wiley Interdiscip. Rev.: Comput.
  Mol. Sci.} \textbf{2011}, \emph{1}, 826--843\relax
\mciteBstWouldAddEndPuncttrue
\mciteSetBstMidEndSepPunct{\mcitedefaultmidpunct}
{\mcitedefaultendpunct}{\mcitedefaultseppunct}\relax
\EndOfBibitem
\bibitem[Marsili et~al.(2006)Marsili, Barducci, Chelli, Procacci, and
  Schettino]{mars+06jpcb}
Marsili,~S.; Barducci,~A.; Chelli,~R.; Procacci,~P.; Schettino,~V.
  \emph{J.~Phys.~Chem.~B} \textbf{2006}, \emph{110}, 14011--14013\relax
\mciteBstWouldAddEndPuncttrue
\mciteSetBstMidEndSepPunct{\mcitedefaultmidpunct}
{\mcitedefaultendpunct}{\mcitedefaultseppunct}\relax
\EndOfBibitem
\bibitem[Barducci et~al.(2008)Barducci, Bussi, and Parrinello]{bard+08prl}
Barducci,~A.; Bussi,~G.; Parrinello,~M. \emph{Phys.~Rev.~Lett.} \textbf{2008},
  \emph{100}, 020603\relax
\mciteBstWouldAddEndPuncttrue
\mciteSetBstMidEndSepPunct{\mcitedefaultmidpunct}
{\mcitedefaultendpunct}{\mcitedefaultseppunct}\relax
\EndOfBibitem
\bibitem[Iannuzzi et~al.(2003)Iannuzzi, Laio, and
  Parrinello]{iann-laio-parr03prl}
Iannuzzi,~M.; Laio,~A.; Parrinello,~M. \emph{Phys.~Rev.~Lett.} \textbf{2003},
  \emph{90}, 238302\relax
\mciteBstWouldAddEndPuncttrue
\mciteSetBstMidEndSepPunct{\mcitedefaultmidpunct}
{\mcitedefaultendpunct}{\mcitedefaultseppunct}\relax
\EndOfBibitem
\bibitem[CPM()]{CPMD}
{CPMD}. \url{http://www.cpmd.org/}, Copyright IBM Corp 1990-2008, Copyright MPI
  f\"ur Festk\"orperforschung Stuttgart 1997-2000\relax
\mciteBstWouldAddEndPuncttrue
\mciteSetBstMidEndSepPunct{\mcitedefaultmidpunct}
{\mcitedefaultendpunct}{\mcitedefaultseppunct}\relax
\EndOfBibitem
\bibitem[Tribello et~al.(2010)Tribello, Ceriotti, and
  Parrinello]{trib-ceri-parr10pnas}
Tribello,~G.~A.; Ceriotti,~M.; Parrinello,~M.
  \emph{Proc.~Natl.~Acad.~Sci.~U.S.A.} \textbf{2010}, \emph{107},
  17509--17414\relax
\mciteBstWouldAddEndPuncttrue
\mciteSetBstMidEndSepPunct{\mcitedefaultmidpunct}
{\mcitedefaultendpunct}{\mcitedefaultseppunct}\relax
\EndOfBibitem
\bibitem[Marto{\v{n}}{\'a}k et~al.(2006)Marto{\v{n}}{\'a}k, Donadio, Oganov,
  and Parrinello]{mart+06natmat}
Marto{\v{n}}{\'a}k,~R.; Donadio,~D.; Oganov,~A.; Parrinello,~M. \emph{Nat.
  Mater.} \textbf{2006}, \emph{5}, 623--6\relax
\mciteBstWouldAddEndPuncttrue
\mciteSetBstMidEndSepPunct{\mcitedefaultmidpunct}
{\mcitedefaultendpunct}{\mcitedefaultseppunct}\relax
\EndOfBibitem
\bibitem[Marto{\v{n}}{\'a}k et~al.(2007)Marto{\v{n}}{\'a}k, Donadio, Oganov,
  and Parrinello]{mart+07prb}
Marto{\v{n}}{\'a}k,~R.; Donadio,~D.; Oganov,~A.; Parrinello,~M.
  \emph{Phys.~Rev.~B} \textbf{2007}, \emph{76}, 14120\relax
\mciteBstWouldAddEndPuncttrue
\mciteSetBstMidEndSepPunct{\mcitedefaultmidpunct}
{\mcitedefaultendpunct}{\mcitedefaultseppunct}\relax
\EndOfBibitem
\bibitem[Best and Hummer(2010)Best, and Hummer]{best-hummer10pnas}
Best,~R.; Hummer,~G. \emph{Proc.~Natl.~Acad.~Sci.~U.S.A.} \textbf{2010},
  \emph{107}, 1088\relax
\mciteBstWouldAddEndPuncttrue
\mciteSetBstMidEndSepPunct{\mcitedefaultmidpunct}
{\mcitedefaultendpunct}{\mcitedefaultseppunct}\relax
\EndOfBibitem
\bibitem[E and Vanden-Eijnden(2004)E, and Vanden-Eijnden]{e-vand04book}
E,~W.; Vanden-Eijnden,~E. In \emph{Multiscale, Modelling, and Simulation};
  Attinger,~S., Koumoutsakos,~P., Eds.; Springer, Berlin, 2004; p 3568\relax
\mciteBstWouldAddEndPuncttrue
\mciteSetBstMidEndSepPunct{\mcitedefaultmidpunct}
{\mcitedefaultendpunct}{\mcitedefaultseppunct}\relax
\EndOfBibitem
\bibitem[Vanden-Eijnden and Tal(2005)Vanden-Eijnden, and Tal]{vand-tal05jcp}
Vanden-Eijnden,~E.; Tal,~F. \emph{J.~Chem.~Phys.} \textbf{2005}, \emph{123},
  184103\relax
\mciteBstWouldAddEndPuncttrue
\mciteSetBstMidEndSepPunct{\mcitedefaultmidpunct}
{\mcitedefaultendpunct}{\mcitedefaultseppunct}\relax
\EndOfBibitem
\bibitem[Hartmann and Sch{\"u}tte(2007)Hartmann, and
  Sch{\"u}tte]{hart-schut07pdnp}
Hartmann,~C.; Sch{\"u}tte,~C. \emph{Phys. D} \textbf{2007}, \emph{228},
  59--63\relax
\mciteBstWouldAddEndPuncttrue
\mciteSetBstMidEndSepPunct{\mcitedefaultmidpunct}
{\mcitedefaultendpunct}{\mcitedefaultseppunct}\relax
\EndOfBibitem
\bibitem[Singh et~al.(2011)Singh, Chiu, and de~Pablo]{sing+11jsp}
Singh,~S.; Chiu,~C.; de~Pablo,~J. \emph{J. Stat. Phys.} \textbf{2011},
  \emph{145}, 932--945\relax
\mciteBstWouldAddEndPuncttrue
\mciteSetBstMidEndSepPunct{\mcitedefaultmidpunct}
{\mcitedefaultendpunct}{\mcitedefaultseppunct}\relax
\EndOfBibitem
\bibitem[Humphrey et~al.(1996)Humphrey, Dalke, and Schulten]{VMD}
Humphrey,~W.; Dalke,~A.; Schulten,~K. \emph{J. Molec. Graphics} \textbf{1996},
  \emph{14}, 33--38\relax
\mciteBstWouldAddEndPuncttrue
\mciteSetBstMidEndSepPunct{\mcitedefaultmidpunct}
{\mcitedefaultendpunct}{\mcitedefaultseppunct}\relax
\EndOfBibitem
\bibitem[Ramachandran et~al.(1963)Ramachandran, Ramakrishnan, and
  Sasisekharan]{ramachandran}
Ramachandran,~G.~N.; Ramakrishnan,~C.; Sasisekharan,~V. \emph{J. Mol. Biol.}
  \textbf{1963}, \emph{7}, 95\relax
\mciteBstWouldAddEndPuncttrue
\mciteSetBstMidEndSepPunct{\mcitedefaultmidpunct}
{\mcitedefaultendpunct}{\mcitedefaultseppunct}\relax
\EndOfBibitem
\bibitem[Kearsley(1989)]{kearsley}
Kearsley,~S.~K. \emph{Acta Cryst. A} \textbf{1989}, \emph{45}, 208--210\relax
\mciteBstWouldAddEndPuncttrue
\mciteSetBstMidEndSepPunct{\mcitedefaultmidpunct}
{\mcitedefaultendpunct}{\mcitedefaultseppunct}\relax
\EndOfBibitem
\bibitem[Branduardi et~al.(2007)Branduardi, Gervasio, and
  Parrinello]{bran+07jcp}
Branduardi,~D.; Gervasio,~F.~L.; Parrinello,~M. \emph{J.~Chem.~Phys.}
  \textbf{2007}, \emph{126}, 054103\relax
\mciteBstWouldAddEndPuncttrue
\mciteSetBstMidEndSepPunct{\mcitedefaultmidpunct}
{\mcitedefaultendpunct}{\mcitedefaultseppunct}\relax
\EndOfBibitem
\bibitem[Raiteri et~al.(2006)Raiteri, Laio, Gervasio, Micheletti, and
  Parrinello]{rait+06jpcb}
Raiteri,~P.; Laio,~A.; Gervasio,~F.~L.; Micheletti,~C.; Parrinello,~M.
  \emph{J.~Phys.~Chem.~B} \textbf{2006}, \emph{110}, 3533--3539\relax
\mciteBstWouldAddEndPuncttrue
\mciteSetBstMidEndSepPunct{\mcitedefaultmidpunct}
{\mcitedefaultendpunct}{\mcitedefaultseppunct}\relax
\EndOfBibitem
\bibitem[Bussi et~al.(2006)Bussi, Gervasio, Laio, and Parrinello]{buss+06jacs}
Bussi,~G.; Gervasio,~F.~L.; Laio,~A.; Parrinello,~M. \emph{J.~Am.~Chem.~Soc.}
  \textbf{2006}, \emph{128}, 13435--13441\relax
\mciteBstWouldAddEndPuncttrue
\mciteSetBstMidEndSepPunct{\mcitedefaultmidpunct}
{\mcitedefaultendpunct}{\mcitedefaultseppunct}\relax
\EndOfBibitem
\bibitem[Piana and Laio(2007)Piana, and Laio]{pian-laio07jpcb}
Piana,~S.; Laio,~A. \emph{J.~Phys.~Chem.~B} \textbf{2007}, \emph{111},
  4553--4559\relax
\mciteBstWouldAddEndPuncttrue
\mciteSetBstMidEndSepPunct{\mcitedefaultmidpunct}
{\mcitedefaultendpunct}{\mcitedefaultseppunct}\relax
\EndOfBibitem
\bibitem[Bonomi. et~al.(2009)Bonomi., Barducci, and Parrinello]{bono+09jcc}
Bonomi.,~M.; Barducci,~A.; Parrinello,~M. \emph{J.~Comp.~Chem.} \textbf{2009},
  \emph{30}, 1615--1621\relax
\mciteBstWouldAddEndPuncttrue
\mciteSetBstMidEndSepPunct{\mcitedefaultmidpunct}
{\mcitedefaultendpunct}{\mcitedefaultseppunct}\relax
\EndOfBibitem
\bibitem[{MacKerell Jr.} et~al.(1998){MacKerell Jr.}, Bashford, Bellot,
  {Dunbrack Jr.}, Evanseck, Field, Fischer, Gao, Guo, Ha, Joseph-McCarthy,
  Kuchnir, Kuczera, Lau, Mattos, Michnick, Ngo, Nguyen, Prodhom, III, Roux,
  Schlenkrich, Smith, Stote, Straub, Watanabe, Wiorkiewicz-Kunczera, Yin, and
  Karplus]{charmm}
{MacKerell Jr.},~A.~D. et~al.  \emph{J. Phys. Chem. B} \textbf{1998},
  \emph{102}, 3586--3616\relax
\mciteBstWouldAddEndPuncttrue
\mciteSetBstMidEndSepPunct{\mcitedefaultmidpunct}
{\mcitedefaultendpunct}{\mcitedefaultseppunct}\relax
\EndOfBibitem
\bibitem[Maragliano et~al.(2006)Maragliano, Fischer, and
  Vanden-Eijnden]{mara-fish-vand06jcp}
Maragliano,~L.; Fischer,~A.; Vanden-Eijnden,~E. \emph{J. Chem. Phys.}
  \textbf{2006}, \emph{125}, 024106\relax
\mciteBstWouldAddEndPuncttrue
\mciteSetBstMidEndSepPunct{\mcitedefaultmidpunct}
{\mcitedefaultendpunct}{\mcitedefaultseppunct}\relax
\EndOfBibitem
\bibitem[Lazaridis et~al.(1991)Lazaridis, Tobias, Brooks, and
  Paulaitis]{dialanine1}
Lazaridis,~T.; Tobias,~D.~J.; Brooks,~C.; Paulaitis,~M.~E. \emph{J. Chem.
  Phys.} \textbf{1991}, \emph{95}, 7612--7625\relax
\mciteBstWouldAddEndPuncttrue
\mciteSetBstMidEndSepPunct{\mcitedefaultmidpunct}
{\mcitedefaultendpunct}{\mcitedefaultseppunct}\relax
\EndOfBibitem
\bibitem[Tobias and Brooks(1992)Tobias, and Brooks]{dialanine2}
Tobias,~D.~J.; Brooks,~C.~L. \emph{J. Phys. Chem.} \textbf{1992}, \emph{96},
  3864--3870\relax
\mciteBstWouldAddEndPuncttrue
\mciteSetBstMidEndSepPunct{\mcitedefaultmidpunct}
{\mcitedefaultendpunct}{\mcitedefaultseppunct}\relax
\EndOfBibitem
\bibitem[Bartels and Karplus(1997)Bartels, and Karplus]{dialanine3}
Bartels,~C.; Karplus,~M. \emph{J. Comput. Chem.} \textbf{1997}, \emph{18},
  1450--1462\relax
\mciteBstWouldAddEndPuncttrue
\mciteSetBstMidEndSepPunct{\mcitedefaultmidpunct}
{\mcitedefaultendpunct}{\mcitedefaultseppunct}\relax
\EndOfBibitem
\bibitem[Apostolakis et~al.(1999)Apostolakis, Ferrara, and
  Caflisch]{dialanine4}
Apostolakis,~J.; Ferrara,~P.; Caflisch,~A. \emph{J. Chem. Phys.} \textbf{1999},
  \emph{110}, 2099--2108\relax
\mciteBstWouldAddEndPuncttrue
\mciteSetBstMidEndSepPunct{\mcitedefaultmidpunct}
{\mcitedefaultendpunct}{\mcitedefaultseppunct}\relax
\EndOfBibitem
\bibitem[Smith(1999)]{dialanine5}
Smith,~P.~E. \emph{J. Chem. Phys.} \textbf{1999}, \emph{111}, 5568--5579\relax
\mciteBstWouldAddEndPuncttrue
\mciteSetBstMidEndSepPunct{\mcitedefaultmidpunct}
{\mcitedefaultendpunct}{\mcitedefaultseppunct}\relax
\EndOfBibitem
\bibitem[Ryckaert et~al.(1977)Ryckaert, Ciccotti, and Berendsen]{shake}
Ryckaert,~J.~P.; Ciccotti,~G.; Berendsen,~H. J.~C. \emph{J. Comput. Phys.}
  \textbf{1977}, \emph{23}, 327--341\relax
\mciteBstWouldAddEndPuncttrue
\mciteSetBstMidEndSepPunct{\mcitedefaultmidpunct}
{\mcitedefaultendpunct}{\mcitedefaultseppunct}\relax
\EndOfBibitem
\bibitem[Phillips et~al.(2005)Phillips, Braun, Wang, Gumbart, Tajkhorshid,
  Villa, Chipot, Skeel, Kal{\'e}, and Schulten]{namd}
Phillips,~J.~C.; Braun,~R.; Wang,~W.; Gumbart,~J.; Tajkhorshid,~E.; Villa,~E.;
  Chipot,~C.; Skeel,~R.~D.; Kal{\'e},~L.; Schulten,~K. \emph{J. Comput. Chem.}
  \textbf{2005}, \emph{26}, 1781--802\relax
\mciteBstWouldAddEndPuncttrue
\mciteSetBstMidEndSepPunct{\mcitedefaultmidpunct}
{\mcitedefaultendpunct}{\mcitedefaultseppunct}\relax
\EndOfBibitem
\end{mcitethebibliography}
\end{document}